\documentclass[12pt]{article}
\usepackage[utf8]{inputenc}
\usepackage{authblk}
\usepackage{setspace}
\usepackage[margin=1.25in]{geometry}
\usepackage{graphicx}
\usepackage{subcaption}
\usepackage{amsmath}
\usepackage{lineno}
\usepackage{lipsum}
\usepackage[comma,square,numbers,sort&compress]{natbib}

\usepackage[dvipsnames]{xcolor}
\usepackage{hyperref}
\hypersetup{                                                          
pdftoolbar    = true,            
pdfmenubar    = true,            
colorlinks    = true,                                                          
linkcolor     = blue,                                                          
citecolor     = blue,                                                       
linktocpage   = true,  
linktocpage   = true,
urlcolor      = CadetBlue,   
}

\newcommand{\ptch}{$p_{\rm{ T, ch jet}}$}
\newcommand{\z}{$z_{||}$}
\newcommand{\zch}{$z_{||}^{\rm{ch}}$}
\newcommand{\sq}{$\sqrt{s}$}

\begin{document}
\title{Investigation of charm-quark fragmentation by correlation and jet measurements with ALICE}
\author[$ $]{Ravindra Singh$^{*}$ {for the ALICE Collaboration}}

\affil[$ $]{Department of Physics, Indian Institute of Technology Indore}
\affil[$ $]{\href{mailto:ravindra.singh@cern.ch}{\texttt{ravindra.singh@cern.ch}}}
\onehalfspacing
\maketitle

\date{}

\begin{abstract}

Measurements of heavy-flavour tagged jets  allow for comparisons of the heavy quarks (charm and beauty) production, propagation, and hadronization across different collision systems. Comparison of measurements performed in pp with Pb--Pb collisions can help in studying the possible modification of the heavy-quark production and hadronization inside jets due to the quark-gluon plasma medium.

This article presents the ALICE measurements of jets tagged with D-mesons in pp collisions at $\sqrt{s}= 5.02$ TeV and $\sqrt{s}= 13$ TeV (including the observation of the dead-cone effect), as well as in Pb--Pb collisions at $\sqrt{s_{\rm{NN}}}= 5.02$ TeV.

\end{abstract}

\newpage
\section{Introduction}

The partons produced in the ultra-relativistic hadronic collisions undergo successive gluon emissions, forming a parton shower where energy is transferred to lower-energy particles, which later hadronize into sprays of the final-state particles known as jets. Measurements of heavy-flavour tagged jets offer a distinct sensitivity to investigate the processes involved in the production of heavy quarks and contributions from higher-order processes like gluon splitting and flavour excitation~\cite{Li:2021gjw}. Such investigations are crucial for testing perturbative quantum chromodynamics (pQCD) calculations and validating Monte Carlo (MC) event generators. Measurements of momentum fraction carried by heavy-flavour hadrons (\z) in jets provide further insights into heavy-quark fragmentation~\cite{ALICE:2019cbr}. In particular, the ALICE measurements offer detailed insights into D$^0$-jet properties as a function of the jet resolution parameter ($R$)~\cite{STAR:2009kkp} and enable the measurement of \zch distributions across a wide range of charged-jet transverse momentum intervals. In heavy-ion collisions, high-energy partons experience interactions with the constituents of the quark-gluon plasma, resulting in gluon radiation and elastic scattering. Differences in quark mass and quark-to-gluon fraction between heavy-flavour and inclusive jets (i.e., jets produced from gluon or light-flavour quark fragmentation) may contribute to different energy loss within the medium~\cite{ALICE:2018lyv}.

In QCD, the parton shower pattern depends on the mass of the emitting parton, due to the presence of the dead-cone effect, which vetoes the gluon radiation in an angular cone with an opening angle $m/E$ with respect to the flight direction of the emitter. Directly measuring this effect is challenging due to hadronization effects, which dilute our access to the QCD parton radiation starting from the set of final-state reconstructed particles. To overcome this limitation, a reclustering technique with the Cambridge-Aachen algorithm is used, in order to obtain access to the QCD splittings~\cite{ALICE:2021aqk}. 

\begin{figure}[h!]
  \centering
    \includegraphics[width=\linewidth]{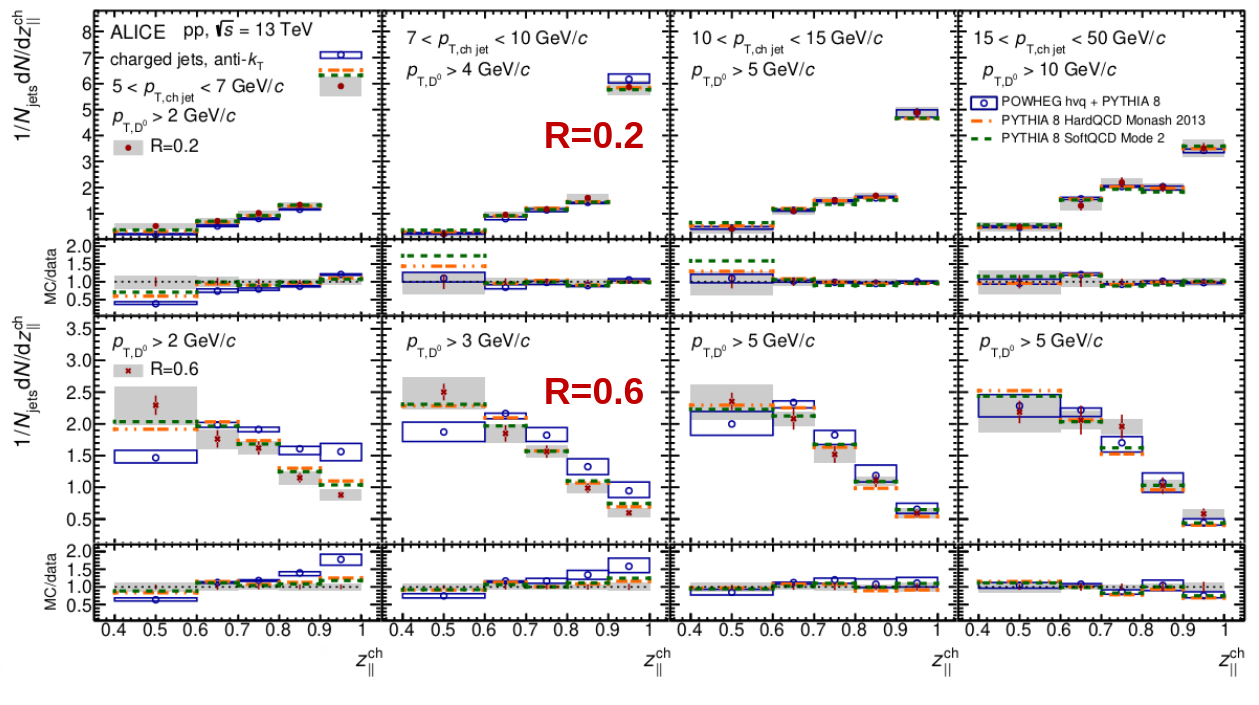}
  \caption{Distributions of \zch differential yield of charm jets tagged with D$^0$-meson normalised by the number of D$^0$ jets in pp collisions at \sq~= 13 TeV.}
  \label{fig:z_dist}
\end{figure}

\begin{figure}[h!]
  \centering
    \includegraphics[width=\linewidth]{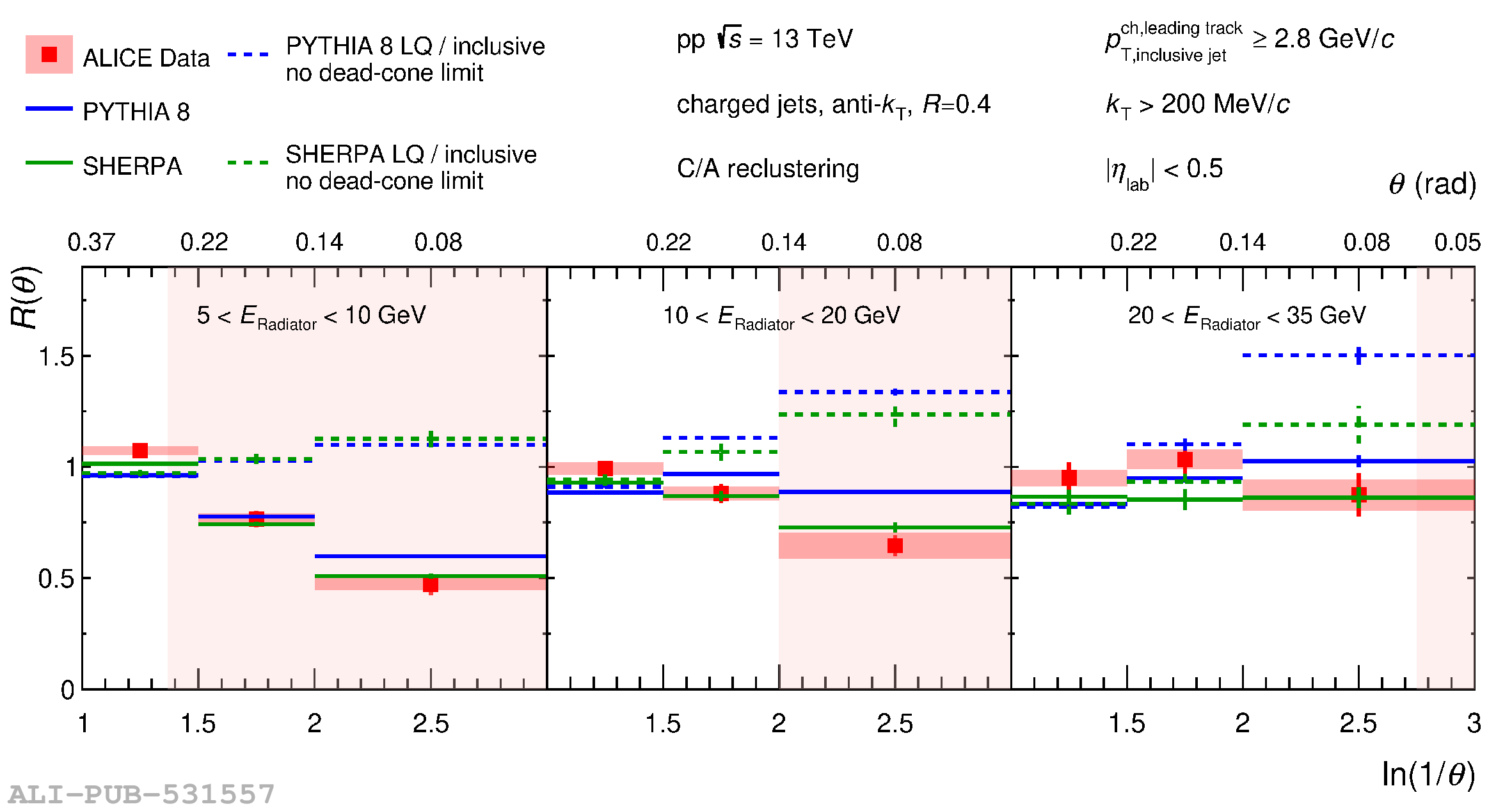}
  \caption{The ratios of the splitting-angle probability distributions for D$^0$-meson tagged jets to inclusive jets, $R(\theta)$, measured in pp collisions at \sq~= 13 TeV.}
  \label{fig:deadcone}
\end{figure}

\begin{figure}[h!]
  \centering
    \includegraphics[width=0.5\linewidth]{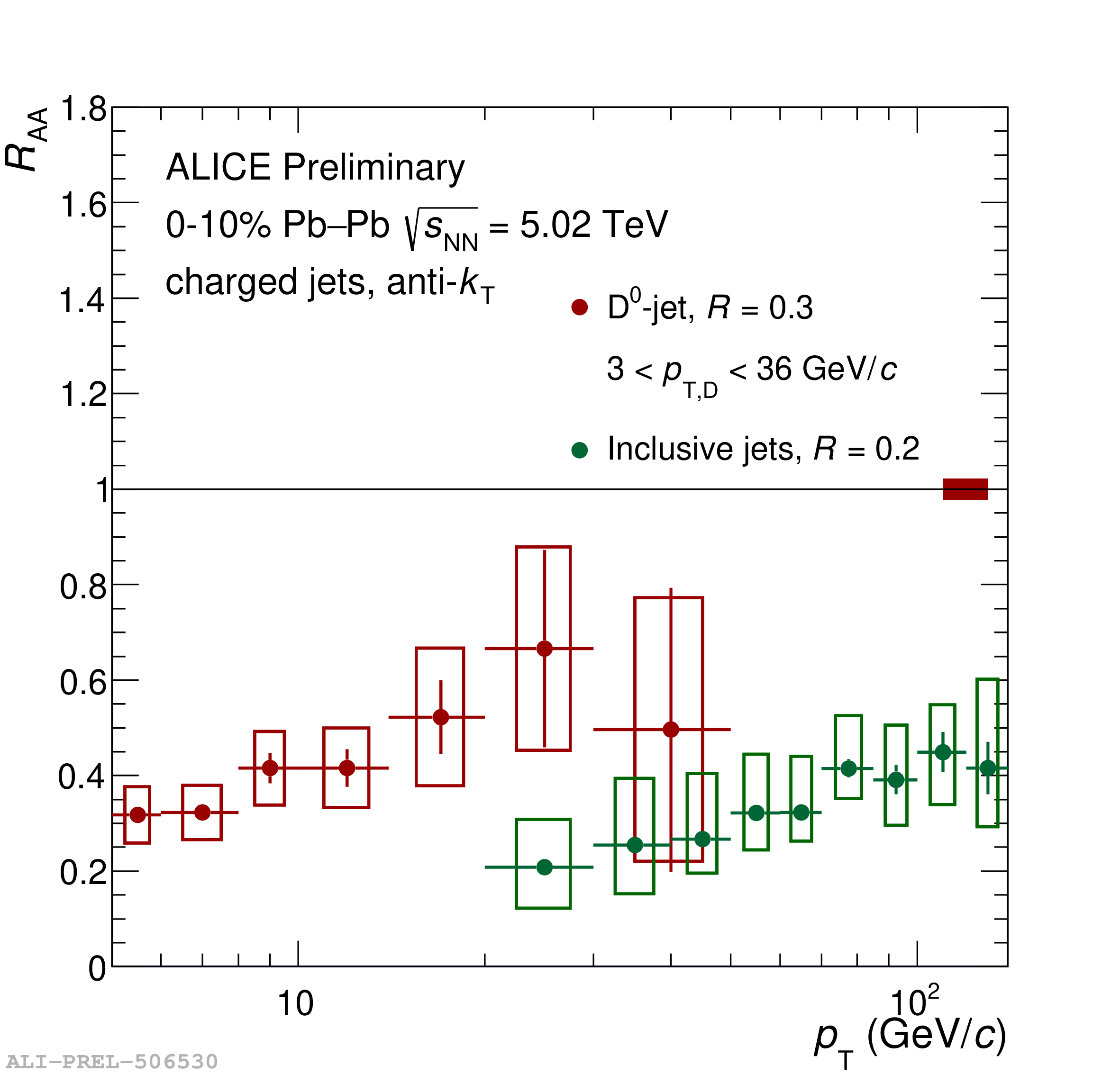}
  \caption{Nuclear modification factor of D$^{0}$-jets (red) and inclusive jets (green) in 0-10\% centrality Pb--Pb collisions.}
  \label{fig:Raa}
\end{figure}

\section{Results and Conclusion}

Fig.~\ref{fig:z_dist} shows the \zch distributions of D$^0$-tagged jets for different \ptch intervals in pp collisions at \sq = 13 TeV~\cite{ALICE:2022mur}. A peak at \zch $\approx$ 1 is observed for D$^0$ jets with 5 $<$ \ptch $<$ 15 GeV$/c$ and $R$ = 0.2, indicating jets consisting solely of the tagged D$^0$ meson. This peak decreases as $R$ and \ptch increase, leading to a softening of fragmentation (\zch). Model predictions from PYTHIA8 with Monash tune and a beyond leading color reconnection (CR-BLC) Mode 2 tune using the SoftQCD process, agree well with the data within uncertainties. However, the data hints at softer fragmentation in the lowest \ptch interval compared to PYTHIA8 predictions. The differences in \zch distribution between the PYTHIA8 Monash and the CR-BLC Mode 2 tunes are minimal. POWHEG + PYTHIA8 predictions agree well with data above \ptch $>$ 10 GeV$/c$, but predict harder fragmentation at lower \ptch, in particular for larger $R$.

Fig.~\ref{fig:deadcone} illustrates the investigation of the dead-cone effect for charm quarks by showing the ratios ($R(\theta)$) of the splitting angle ($\theta$) distributions for D$^0$-meson tagged jets and inclusive jets, in bins of $E_{\mathrm{radiator}}$, as a function of $\ln{(1/\theta)}$~\cite{ALICE:2021aqk}. The results reveal a suppression of emissions at small angles for charm-tagged jets compared to inclusive jets, with a larger degree of suppression observed at lower radiator energy, in agreement with the expectations in the presence of a dead-cone effect. Monte Carlo simulations performed with PYTHIA8 and SHERPA, including the dead-cone effect in the parton showering, reproduce the data well, differently from cases where a no-dead-cone condition is mimicked by considering massless quarks in place of D0-jets in the numerator of $R(\theta)$. These comparisons support the conclusion that the observed suppression is consistent with the impact of the dead-cone effect associated with the mass of the quarks.

In order to investigate the medium effects on the charm quarks propagating in the QGP, the $R_{\rm{AA}}$ of charm jets tagged by the presence of D$^0$ mesons is shown in Fig.~\ref{fig:Raa} in red markers and compared with the same observable obtained on a sample of inclusive jets~\cite{ALICE:2018lyv,ALICE:2019nrq}. The inclusive jets $R_{\rm{AA}}$ shows a hint of higher suppression than the D$^0$-jet $R_{\rm{AA}}$ in central (0-10\%) Pb--Pb collisions with a significance of $\approx$ 2.1$\sigma$ in the transverse momentum range 20 $<$ \ptch $<$ 30 GeV$/c$ and 1.8$\sigma$ across the entire transverse momentum region where both measurements are overlapped.

In summary, the results presented in this contribution demonstrate an observed softening of fragmentation, and the presence of the dead-cone effect, in charm-tagged jets, as well as a possible reduction in the suppression of charm jets tagged with D$^0$ mesons compared to inclusive jets in Pb--Pb collisions.

\end{document}